\documentclass[twocolumn, pra, showpacs,superscriptaddress]{revtex4}
\usepackage{graphicx}
\usepackage{amsmath}
\usepackage{bm}
\usepackage{float}
\usepackage{color}
\usepackage{sidecap}
\usepackage{amsfonts}
\usepackage{bbold}
\usepackage{dcolumn}
\usepackage{multirow}

\begin{document}
\title{Gauge-spin-space rotation invariant vortices \\ in spin-orbit coupled Bose-Einstein condensates}
\author{Zhi-Fang Xu}
\affiliation{Department of Physics, University of Tokyo, 7-3-1 Hongo, Bunkyo-ku, Tokyo 113-0033, Japan}
\author{Shingo Kobayashi}
\affiliation{Department of Physics, University of Tokyo, 7-3-1 Hongo, Bunkyo-ku, Tokyo 113-0033, Japan}
\author{Masahito Ueda}
\affiliation{Department of Physics, University of Tokyo, 7-3-1 Hongo, Bunkyo-ku, Tokyo 113-0033, Japan}

\date{\today}

\begin{abstract}
  We revisit ground states of spinor Bose-Einstein condensates with a Rashba spin-orbit coupling,
and find that votices show up as a direct consequence of spontaneous symmetry breaking into a combined gauge, spin,
and space rotation symmetry, which determines the vortex-core spin state at the rotating center.
For the continuous combined symmetry, the total spin rotation about the rotating axis is restricted to $2\pi$,
whereas for the discrete combined symmetry, we further need $2F$ quantum numbers to characterize 
the total spin rotation for the spin-$F$ system.
For lattice phases we find that in the ground state the topological charge for each unit cell vanishes.
However, we find two types of highly symmetric lattices with a nontrivial topological charge in the spin-$\frac{1}{2}$ system based on the symmetry classification,
and show that they are skyrmion crystals. 
\end{abstract}


\maketitle


\section{Introduction}

Vortices are ubiquitous in a rich variety of systems 
from Bose-Einstein condensates (BECs) \cite{yuki2012,stamper-kurn2012},
superfluid helium \cite{salomaa1987} to superconductors \cite{blatter1994}.
In scalar BECs and superfluid helium-4, vortices are quantized, in units of a topological invariant called mass circulation
$\oint_{\Gamma}\mathbf{v}^{\rm mass}\cdot d\bm{\ell}=2\pi\hbar n_{v}/M$, where $\Gamma$ is a close
loop enclosing the vortex, $\mathbf{v}^{\rm mass}$ is the superfluid velocity, and $n_{v}$ is an integer.
When internal degrees of freedom are involved, topologically stable fractional vortices
emerge in the polar phase and the cyclic phase of spinor BECs due to the underlying $\mathbb{Z}_2$ \cite{zhou2001, yuki2012}
and tetrahedral symmetries \cite{semenoff2007}, respectively. 
To classify or predict these and other topological excitations
in spinor BECs such as monopoles, skyrmions and knots, we can invoke the homotopy theory \cite{mermin1979},
because the spin and orbital degrees of freedom are decoupled.

To define topological invariants to characterize quantized vortices 
and the quantization of vortices, we usually require 
that the order parameter at the boundary belongs to a prescribed order parameter manifold of a certain topological nature.
However, knowing the boundary of the vortex is, in general, insufficient to predict the order parameter
close to the vortex core, because vortices having the same topological number can have different cores \cite{kobayashi2012,kobayashi2009}.
Therefore, to distinguish vortices with different types of cores, we should resort to other methods.
One such method is presented by Salomaa and Volovik 
in Refs. \cite{salomaa1983,salomaa1986}, where they applied a symmetry classification scheme to distinguish
different vortices within the same topological class of the superfluid $^{3}$He B-phase
under an axisymmetric boundary condition, according to the axisymmetry and three discrete internal symmetries.

Here, we show that the symmetry classification scheme 
can also be applied to distinguish vortices in BECs in which the spin and orbital
degrees of freedom are coupled.
For spin-orbit (SO) coupled spinor BECs \cite{lin2011,zhang2012}
and spinor dipolar BECs \cite{griesmaier2005,vengalattore2008,pasquiou2011},
the spin and orbital degrees of freedom are coupled due to the single-particle SO coupling or the long-range dipolar interaction, 
and therefore vortices can spontaneously emerge in the ground states.
Examples are two different vortices found in ground states of a spinor dipolar BEC \cite{yuki2006,yi2006}, with their cores 
filled by the polar phase or the ferromagnetic phase.
For a SO-coupled spinor BEC, ground states 
with a single vortex or vortex lattices have been found \cite{wu2011,xu2011,kawakami2011,hu2012,sinha2011,xu2012,ruokokoski2012}. 

In this article, we systematically investigate
vortices in the ground states of a SO-coupled spinor BEC from the point of view of spontaneous symmetry breaking (SSB)
and find that they are direct consquences of SSB into a combined gauge, spin, and space rotation symmetry,
as the system is invariant under the U(1) gauge transformation, the combined spin and space rotation
and the time reversal. 
The vortex-core spin state is then determined by the spin-gauge symmetry.
As long as the spin and orbital degrees of freedom are coupled, possible combined symmetries are reduced,
because the spin and space must be rotated simultaneously to make the Hamiltonian invariant.
For lattice phases found in the ground state, vortices with different combined symmetries form lattice structures, 
and the topological charge for each unit cell vanishes. However, in a SO-coupld spin-$\frac{1}{2}$ BEC, we find highly-symmetric skyrmion crystals, 
which are constructed from superposition of several plane waves.

Our symmetry classification method differs from that in Refs.~\cite{salomaa1983, salomaa1986} in the following respects.
In Refs.~\cite{salomaa1983, salomaa1986}, the 
authors consider vortices in the superfluid $^3$He-B within the same topological class by assuming 
an axisymmetric boundary condition.
For SO-coupled BECs, we do not take such a boundary constraint.
If a vortex preserves the continuous combined symmetry,
the total spin rotation about the rotating axis is $2\pi$.
However, if the vortex is invariant under the discrete combined symmetry,
the symmetry alone is not enough to determine the total gauge and spin-state rotations 
about the rotating axis, and we need one additional quantum number to count the total gauge rotation and $2F$ quantum numbers to characterize
the total spin rotation for the spin-$F$ system. 
Therefore, vortices preserving the continuous and discrete combined symmetries are not
always fall into the same topological class.
From numerical results shown below, we find vortices preserving the discrete combined symmetry
and cannot be adiabatically deformed into vortices which are invariant under the continuous combined gauge, spin, and space rotation.

This paper is organized as follows. Section II describes the model Hamiltonian for SO-coupled spinor BECs.
Section III classifies vortices from their combined symmetries and determines the corresponding vortex-core spin state 
at the rotating center. Section IV discusses different types of vortices which are invariant under the continous and discrete
combined symmetries found in SO-coupled spinor BECs.
Section V extends the symmetry classification of vortices into spinor BECs.
Section VI summarizes the main results.

\section{Model Hamiltonian}
\label{hamiltonian}
We consider a quasi-two-dimensional spin-$F$ BEC with $N$ atoms and a Rashba SO coupling.
The effective Hamiltonian is $\mathcal{H}=\mathcal{H}_0+\mathcal{H}_{\rm int}$. Here,
the single-particle Hamiltonian is given by
\begin{eqnarray}
  \mathcal{H}_0=\int d\bm{\rho}\, \hat{\psi}^{\dag}\left[
    \frac{\mathbf{p}^2}{2M}+\mathcal{V}_o+\frac{v}{F}(p_xF_x+p_yF_y)\right]\hat{\psi},
  \label{singleparticle}
\end{eqnarray}
where $\hat{\psi}=(\hat{\psi}_F,\dots,\hat{\psi}_{-F})^T$, $\bm{\rho}\equiv(x,y)$, and
$M$, $\mathcal{V}_o=M\omega_{\perp}^2\bm{\rho}^2/2$, and $F_{x,y}$ are the atomic mass,
the trapping potential, and the spin-$F$ matrices, respectively.
The interaction part of the Hamiltonian depends on the atomic spin and it is assumed to take the same form as that described in Ref. \cite{xu2012}, where
\begin{eqnarray}
  \mathcal{H}_{\rm int}&=&\frac{1}{2}\int d\bm{\rho}\left(\alpha \hat{\psi}_i^{\dag}\hat{\psi}_j^{\dag}\hat{\psi}_j\hat{\psi}_i
  +\beta \hat{A}+\gamma\hat{B}\right), \\
  \hat{A}&=&
     \begin{cases}
       :(\hat{\psi}_{\frac{1}{2}}^{\dag}\hat{\psi}_{\frac{1}{2}}-\hat{\psi}_{-\frac{1}{2}}^{\dag}\hat{\psi}_{-\frac{1}{2}})^2:,
       & \text{if $F=\frac{1}{2}$,} \\
       \hat{\psi}_i^{\dag}\hat{\psi}_k^{\dag}\vec{F}_{ij}\cdot\vec{F}_{kl} \hat{\psi}_l\hat{\psi}_j,
       & \text{if $F=1,2$,} 
     \end{cases}
  \\
  \hat{B}&=&(-1)^{i+j}\hat{\psi}_i^{\dag}\hat{\psi}^{\dag}_{-i} \hat{\psi}_j\hat{\psi}_{-j},\quad \text{if $F=2$}. 
  \label{interaction}
\end{eqnarray}
Because the singlet-pairing interaction arises only for the spin-2 case,
we choose $\gamma=0$ for the spin-$\frac{1}{2}$ and spin-1 cases.

The above Hamiltonian possesses the full symmetry $\rm G=U(1)\times SO(2)\times \mathcal{T}$,
which describes the U(1) gauge transformation, the SO(2) simultaneous spin and space rotation about the $z$-axis,
and the time reversal, respectively. According to the symmetry classification scheme \cite{volovik1985,bruder1986,yip2007,yuki2011}, 
we classify ground states by their remaining symmetry described by the group isotropy H \cite{xu2012}.
When the state preserves the continuous or discrete combined gauge, spin, and space rotations,
usually a vortex appears at the rotating center, with the vortex core completely or partially determined 
by the corresponding spin-gauge symmetry. This is similar to but different from the situation discussed in Ref. \cite{volovik1985},
where the combined gauge and space rotation symmetry plays an important role in determining zeros of
the superconducting order parameter of heavy-fermion systems; in the present case, there are spin degrees of freedom
and vortices can be nonsingular for SO-coupled BECs.

\section{Symmetry classification of vortices} 
\label{symmetry}

\subsection{Salomaa and Volovik's method}
In this subsection, we briefly review the symmetry
classification scheme on vortices in the superfluid $^{3}$He-B phase by Salomaa and Volovik
\cite{salomaa1983,salomaa1986}.
The procedure of this scheme is as follows:

(1) Firstly, an axisymmetric boundary condition is assumed. Far away from the vortex core, the order parameter
is assumed to be $A_{\alpha j}(\rho=\infty,\varphi,z)=CR_{\alpha j}e^{im\varphi}$, where
$\rho$, $\varphi$, and $z$ denote cylindrical coordinates and $R_{\alpha j}$ describes the relative rotation
of the spin ($\alpha$) and orbit ($j$). By approximately choosing the spin coordinates, we can set $R_{\alpha j}=\delta_{\alpha j}$.
Therefore, the operator describing the axisymmetry is given by $\hat{Q}=\hat{L_z}+\hat{S}_z-m\hat{I}$,
where $\hat{L}_z=-i\partial/\partial\varphi+\hat{L}_z^{\rm int}$ is the total operator of the $z$-component orbital angular momentum,
including the external and internal ($\hat{L}_z^{\rm int}$) orbital rotations,
$\hat{S}_z$ is the $z$-component of the spin operator, and $\hat{I}$ is the identity operator.

(2) Secondly, if vortices preserve the continuous combined gauge, spin, and space rotation symmetry,
the order parameter should satisfy $\hat{Q}A_{\alpha j}=0$. This equation is met if $A_{\alpha j}$ takes the following form:
\begin{eqnarray}
  A_{\alpha j}=\sum\limits_{\mu\nu}C_{\mu\nu}(\rho,\varphi,z)\lambda_{\alpha}^{\mu}\lambda_j^{\nu}e^{i(m-\mu-\nu)\varphi},
  \label{he3B}
\end{eqnarray}
where $\lambda_{j}^{\nu}$ and $\lambda_{\alpha}^{\mu}$ are eigenfuctions of operators $\hat{L}_z^{\rm int}$ and $\hat{S}_z$
with the correponding eigenvalues $\nu$ and $\mu$, respectively.
From Eq.~(\ref{he3B}), we can infer that if $m-\mu-\nu\ne0$, $C_{\mu\nu}(\rho=0)=0$.
Therefore, the continuous combined symmetry reduces the allowed vortex-core state.
However, when different values of $\mu$ and $\nu$ all satisfy $m-\mu-\nu$,
the vortex core cannot be uniquely determined.
In Ref.~\cite{salomaa1983}, Salomaa and Volovik consider three discrete internal symmetries $\mathcal{P}_1$, $\mathcal{P}_2$ and $\mathcal{P}_3$ to 
further reduce the ambiguity of the vortex core, with 
\begin{eqnarray}
  \mathcal{P}_1=\mathcal{P}e^{im\pi},\quad \mathcal{P}_2=\mathcal{T}L_{y,\pi}S_{y,\pi},\quad, \mathcal{P}_3=\mathcal{P}_1\mathcal{P}_2,
\end{eqnarray}
where $\mathcal{P}$ is the parity transformation $\vec{\rho}\rightarrow-\vec{\rho}$, $\mathcal{P}_1$ 
denotes the combined gauge and the $\pi$ space rotation about the $z$-axis,
and $\mathcal{P}_2$ denotes the combined time reversal, the $\pi$ space rotation ($L_{y,\pi}$) and $\pi$ spin rotation ($S_{y,\pi}$) about
the $y$-axis.

(3) Thirdly, Salomaa and Volovik also consider vortices break the axisymmetry under the same axisymmetric boundary condition as above.
To make a vortex break axisymmetry, they superpose eigenfunctions of the operator $\hat{Q}$ having different eigenvalues.
For example, if the order parameter takes the form $A_{\alpha j}=A_{\alpha j}(Q=0)+A_{\alpha j}(Q=2)$, where 
\begin{eqnarray}
  &\hat{Q}A_{\alpha j}(Q)=QA_{\alpha j}(Q),& \nonumber\\
  &A_{\alpha j}(Q)=\sum\limits_{\mu\nu}C_{\mu\nu}(Q)\lambda_{\alpha}^{\mu}\lambda_j^{\nu}e^{iQ\varphi}e^{i(m-\mu-\nu)\varphi},&
\end{eqnarray}
and $C_{\mu\nu}(Q\ne0,\rho=\infty)=0$, the corresponding vortex is invariant under $e^{iQ\pi}$.
However, as the boundary condition is fixed, the phase winding number for each component is fixed to be $m-\mu-\nu$
and this vortex is topologically equivalent to the one preserving the continuous combined symmetry.

\subsection{Our method from a SSB point of view}
\label{ourmethod}
In this subsection, we systematically classify vortices in SO-coupled spinor BECs from a SSB point of view.
When the space rotation is involved in the SSB, a vortex might emerge.
In general, to classify the symmetry of vortices, we should consider different subgroups of 
the full symmetry group G of the Hamiltonian.
Due to the SO coupling, we need to simultaneously rotate the same angle
of the spin and the space, which therefore simplifies the procedure to find subgroups of G.

Firstly, we neglect the time-reversal symmetry. Then, there are two different remining symmetries,
that is (1) continuous symmetry and (2) discrete symmetry. The corresponding generator 
is given by $e^{i\phi}\mathcal{C}_z(\theta)$, where $\mathcal{C}_z(\theta)$ refers to the combined
$\theta$ spin-space rotation about the $z$-axis. For the states with continuous symmetry we have 
$\theta\in[0,2\pi)$ and $\phi=\kappa\theta$ with $\kappa=0,\pm1, \pm2,\dots$ ($\kappa=\pm1/2,\pm3/2,\dots)$
for the integer (half-integer) spin case,
whereas for the states with discrete symmetry we have 
$\theta=2\pi/n$ ($n\in\mathbb{Z}$) and $\phi=\kappa\theta$ with $\kappa=0,\dots,n-1$ ($\kappa=1/2,\dots,n-1/2)$
for the integer (half-integer) spin case.

To preserve the combined gauge, spin, and space rotation symmetries, the vortex-core spin state $|\psi\rangle$ located at the rotating center must satisfy
the following spin-gauge symmetry:
\begin{eqnarray}
  e^{i\kappa\theta}e^{-iF_z\theta}|\psi\rangle=|\psi\rangle.
  \label{spingauge}
\end{eqnarray}

We first focus on the continuous symmetry. 
In constrast to the superfluid helium-3, where the continuous combined symmetry
is not usually sufficient to uniquely determine the vortex core, 
the order parameter of a SO-coupled BEC is described by $2F+1$ complex numbers and we can uniquely determine the state of the vortex core
as (1) nonsingular with $|\psi\rangle=|M_F=\kappa\rangle$ if $-F\le\kappa\le F$  and (2) singular with the vacuum core
if $|\kappa|>F$. For instance, in the $F=1$ case, two different nonsingular vortices correspond to the vortex core filled by 
the polar and the ferromagnetic phases. 

\begin{figure}[tpb]
\centering
\includegraphics[width=3.0in]{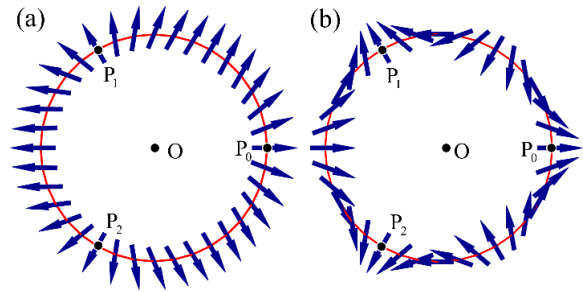}
\caption{(color online). Schematic spin textures for two different vortices in a spin-$\frac{1}{2}$ system
  preserving the same discrete combined symmetry described by $e^{i\kappa 2\pi/3}\mathcal{C}_z(2\pi/3)$,
  where the total spin rotation about the rotating axis at O is $2\pi$ for (a) and $-4\pi$ for (b).  }
\label{fig1}
\end{figure}

\begin{table*}
\centering
\caption{Combined symmetry, vortex generator, and vortex-core state. For the discrete combined symmetry,
  we consider $n\ge3$ for the spin-$\frac{1}{2}$ and spin-1 cases, and $n=3, 4$ for the spin-2 case. For the case of $n=2$, we
  cannot predict whether there is a vortex. When no spin state
  at rotating center can satisfy the corresponding spin-gauge symmetry of Eq. (\ref{spingauge}), singular vortices should appear with the vortex core filled
by the vacuum/normal state.}
\begin{tabular}[b]{c|c|c}
\hline\hline
combined symmetry & vortex generator  & vortex-core state \\ \hline
\multirow{2}{*}{continous} & \multirow{2}{*}{$e^{i\kappa\theta}\mathcal{C}_z(\theta)$ with $\theta\in[0,2\pi)$} 
& $|M_F=\kappa\rangle$ if $\kappa=-F,\dots, F$ \\ & & vacuum/normal state if $|\kappa|>F$  \\ \hline 
\multirow{3}{*}{discrete ($F=\frac{1}{2}$)} 
  & $e^{i\pi/n}\mathcal{C}_{z}(2\pi/n)$  & spin-up \\ 
  & $e^{i(2n-1)\pi/n}\mathcal{C}_{z}(2\pi/n)$  & spin-down \\ 
  & $e^{i\kappa 2\pi/n}\mathcal{C}_{z}(2\pi/n)$ with $\kappa=3/2,\dots, n-3/2$ & vacuum/normal state \\ \hline
\multirow{4}{*}{discrete ($F=1$)} 
  & $e^{i2\pi/n}\mathcal{C}_{z}(2\pi/n)$  & $|M_F=1\rangle$ (ferromagnetic) \\ 
  & $\mathcal{C}_{z}(2\pi/n)$  & $|M_F=0\rangle$ (polar) \\ 
  & $e^{i(n-1)2\pi/n}\mathcal{C}_{z}(2\pi/n)$  & $|M_F=-1\rangle$ (ferromagnetic) \\ 
  & $e^{i\kappa 2\pi/n}\mathcal{C}_{z}(2\pi/n)$ with $\kappa=2,\dots,n-2$ & vacuum/normal state \\ \hline
\multirow{6}{*}{discrete ($F=2$)} 
  & $\mathcal{C}_{z}(2\pi/n)$  & $|M_F=0\rangle$ (uniaxial nematic) \\ 
  & $e^{i2\pi/3}\mathcal{C}_{z}(2\pi/3)$  & arbitrary superposition of $|M_F=1\rangle$ and $|M_F=-2\rangle$ \\ 
  & $e^{i4\pi/3}\mathcal{C}_{z}(2\pi/3)$  & arbitrary superposition of $|M_F=2\rangle$ and $|M_F=-1\rangle$ \\ 
  & $e^{i\pi/2}\mathcal{C}_{z}(\pi/2)$  & $|M_F=1\rangle$ (ferromagnetic) \\ 
  & $e^{i3\pi/2}\mathcal{C}_{z}(\pi/2)$  & $|M_F=-1\rangle$ (ferromagnetic) \\ 
  & $e^{i\pi}\mathcal{C}_{z}(\pi/2)$ and $\mathcal{T}$ & $(e^{-i\phi/2},0,0,0,e^{i\phi/2})^T$ with $\phi$ arbitrary (biaxial nematic) \\ \hline\hline
\end{tabular}
\label{tabl}
\end{table*}

Next, we turn to the discrete symmetry case.
Different from the continuous symmetry case, whether we can uniquely determine the core state depends on the values of
$\kappa$, $n$ and $F$. For instance, when $\kappa=0$, there is only one type of nonsingular vortex core with $|\psi\rangle=|M_F=0\rangle$
for the integer spin case, whereas when $\kappa=1$, $n=3$, and $F=2$, we can find that an arbitrary superposition of 
$|M_F=1\rangle$ and $|M_F=-2\rangle$ satisfies Eq. (\ref{spingauge}), which means that we cannot uniquely
determine the nonsingular vortex-core state. 
Table \ref{tabl} summarizes the relation between the combined symmetry
and the vorte-core spin state. 

No matter whether we can uniquely determine the vortex-core spin state for the discrete combined symmetry,
there are still ambiguities for the total gauge and spin rotations around the vortex core,
which is different from the case of continuous combined symmetry, where the total spin rotation about the rotating
axis is $2\pi$ and the total gauge rotation is $2\kappa\pi$.
Figure~\ref{fig1} schematically illustrates two different vortices in a spin-$\frac{1}{2}$ system which preserve the same discrete combined symmetry
described by $e^{i\kappa 2\pi/3}\mathcal{C}_z(2\pi/3)$.
From this form, we can infer that the allowed total spin rotation along the loop $P_0P_1$ 
is $2\pi/3+2\pi \mathbb{N}_s$ ($\mathbb{N}_s\in\mathbb{Z}$) with (a) $\mathbb{N}_s=0$ and (b) $\mathbb{N}_s=-1$.
A similar ambiguity exists for the gauge rotation.
Therefore, to remove the ambiguity, we can define quantum numbers 
to characterize vortices preserving the same discrete combined symmetry.

For the spin-$\frac{1}{2}$ case, two quantum numbers $\mathbb{N}_g$ and $\mathbb{N}_s$
are needed to remove the ambiguity in the gauge and spin rotations, respectively.
If the  vortice is invaraint under $e^{i\kappa 2\pi/n}\mathcal{C}_z(2\pi/n)$, 
the total gauge rotation and spin rotation about the rotating axis are
$2\kappa\pi+2\pi n\mathbb{N}_g$ and $2\pi+2\pi n\mathbb{N}_s$, respectively,
where $\mathbb{N}_g-\bmod(\mathbb{N}_s,2)/2\in\mathbb{Z}$,
because if we rotate the half-integer spin for $2\pi$, we need to perform a $\pi$ gauge rotation to make the state invariant.
To determine two quantum numbers from numerical results, we can focus
on the phase winding number for the spin component $|M_F\rangle$ defined as
\begin{eqnarray}
  w_{M_F}=\frac{1}{2\pi}\oint_{\Gamma}\frac{\psi_{M_F}^*\nabla \psi_{M_F}-(\nabla\psi_{M_F}^*)\psi_{M_F}}{2|\psi_{M_F}|^2i}\cdot d\vec{\ell},
  \label{windingnumber}
\end{eqnarray}
where $\Gamma$ is a loop around the vortex core, as we have
\begin{eqnarray}
  w_{M_F}=\kappa+n\mathbb{N}_g-(1+n\mathbb{N}_s)M_F.
  \label{windingnumber2}
\end{eqnarray}
If $\mathbb{N}_s\ne0$, we find a topologically different vortex in comparison with the vortex
which preserves the continuous combined rotation symmetry.

Different from the spin-$\frac{1}{2}$ case, for the integer spin-$F$ case, we need $2F+1$ quantum numbers: one quantum number $\mathbb{N}_g$ for 
describing the total gauge rotation and $2F$ quantum numbers for denoting the 
total spin rotation, because we need $2F$ vectors for each spin state in the
Majorana representation \cite{yuki2011,barnett2006}.
Equivalently, we can define another $2F$ quantum numbers $\mathbb{N}_s^{(M_F)}$ for removing the ambiguity of the total spin rotation
by evaluating phase winding numbers for nonzero spin components. 
If the vortex is invariant under $e^{i\kappa 2\pi/n}\mathcal{C}_z(2\pi/n)$, we have
\begin{eqnarray}
  &w_{M_F}=\kappa+n\mathbb{N}_g-\left(1+n\mathbb{N}^{(M_F)}_s\right)M_F,\quad \text{if $M_F\ne0$},&\nonumber\\
  &w_{0}=\kappa+n\mathbb{N}_g,&
  \label{windingnumberF}
\end{eqnarray}
where $\mathbb{N}_g\in\mathbb{Z}$ and $\mathbb{N}_s^{(M_F)}\times M_F\in\mathbb{Z}$.
If at least one quantum number $\mathbb{N}_s^{(M_F)}$ is nonzero, 
the phase winding numbers for all spin components cannot be 
written in a compact form as $w_{M_F}=-M_F+\mathbb{N}_c$, with $\mathbb{N}_c\in\mathbb{Z}$,
which is the salient difference compared with
vortices preserving the continuous combined rotation symmetry.

Taking into account the time-reversal symmetry, similar arguments can be applied. In special cases,
the allowed vortex-core spin state is further restricted due to the time-reversal symmetry.
For instance, for the spin-2 case, if a vortex preserves the discrete combined symmetry
described by $e^{i\pi}\mathcal{C}_z(\pi/2)$, the vortex-core spin state can takes an arbitrary superposition of $|M_F=2\rangle$
and $|M_F=-2\rangle$. If we require the vortex be invariant under the time reversal,
the allowed vortex-core spin states are further reduced, as listed in Table~\ref{tabl}.

\section{Vortices in spin-orbit-coupled BECs}
In this section, we show how to understand different vortices found in the ground states of SO-coupled BECs from the standpoint of combined symmetries.

\subsection{Continuous combined symmetry} 

\begin{figure*}[t]
\centering
\includegraphics[width=2\columnwidth]{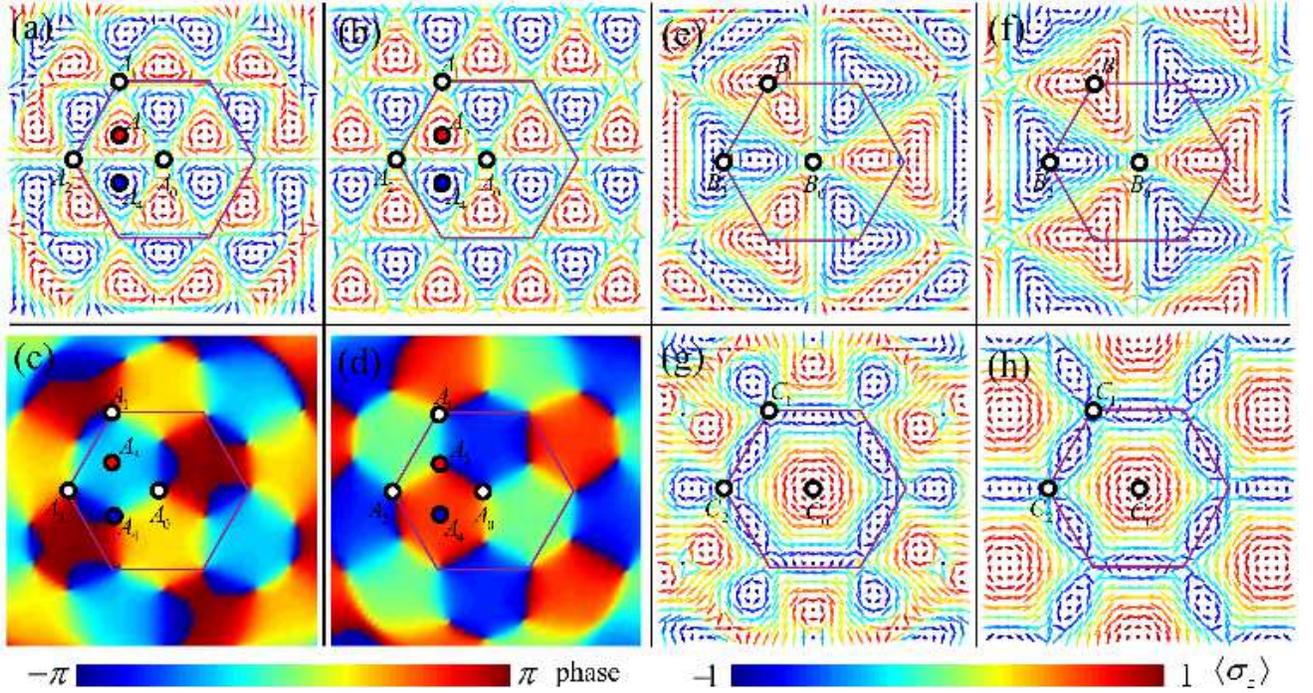}
\caption{(color online). Ground-state spin textures
for three different lattice phases showing triangular (a)
and kagaome (e, g) structures for a spin-orbit-coupled pseudo-spin-$\frac{1}{2}$ BEC, 
with $v'=15$ and (a) $\alpha=0.5\hbar\omega_{\perp}/N$ and $\beta=-0.1\alpha$, 
(e) $\alpha=0.3\hbar\omega_{\perp}/N$ and $\beta=0.1\alpha$, and (g)
$\alpha=0.4\hbar\omega_{\perp}/N$ and $\beta=0.1\alpha$.
The corresponding spin textures for the variational order parameters constructed by a superposition of 
several plane waves given in Eq.~(\ref{latticephases}) are shown in (b), (f), and (h), respectively.
Here the spin polarizations along the $x$-$y$ plane are denoted by arrows and 
its $z$-component $\langle\sigma_z\rangle$ is shown according to the color gauge on the right bottom.
The phase distributions of (c) spin-up and (d) spin-down components for the ground state of the triangular-lattice phase in (a)
are also shown according the color gauge on the left bottom.
The size of each figure is $[-a_{\perp}/2,a_{\perp}/2]\times[-a_{\perp}/2,a_{\perp}/2]$. 
A unit cell of the order parameter is indicated by a solid hexagon.
Singular vortices indicated by open circles form a triangular lattice for all cases.
Two nonsingular vortices are indicated by filled circles with blue ($A_3$) and red ($A_4$) colors. }
\label{fig2}
\end{figure*}

We first focus on the vortices invariant under the continuous combined gauge, spin, and space rotation symmetries. 
In Refs. \cite{xu2012}, approximate low-lying single-particle eigenstates
of the Hamiltonian $\mathcal{H}_0$ with strong SO coupling are presented in momentum space as 
\begin{eqnarray}
  \psi_{\bar{n},m}(\mathbf{k})&\propto& k'^{-1/2}e^{-(k'-v')^2/2}H_{\bar{n}}(k'-v')\nonumber\\
  &&\times e^{im\varphi_k}\zeta_{-F}(\varphi_k),
  \label{eigenstates}
\end{eqnarray}
where $\mathbf{k}$ is the wave vector, $\varphi_k=\arg(k_x+ik_y)$, $a_{\perp}=\sqrt{\hbar/M\omega_{\perp}}$, 
$v'=v/\omega_{\perp}a_{\perp}$, $k'=ka_{\perp}$, $H_{\bar{n}}$ are Hermite polynomials and 
\begin{eqnarray}
  &\zeta_{-\frac{1}{2}}(\varphi_k)=(1,-e^{i\varphi_k})^T/\sqrt{2},&\nonumber\\
  &\zeta_{-1}(\varphi_k)=(e^{-i\varphi_k},-\sqrt{2},e^{i\varphi_k})^T/2,&\nonumber\\
  &\zeta_{-2}(\varphi_k)=(e^{-2i\varphi_k},-2e^{-i\varphi_k},\sqrt{6},-2e^{i\varphi_k},
  e^{2i\varphi_k})^T/4.&\quad
  \label{planewaves}
\end{eqnarray}
By the Fourier transformation, we obtain the corresponding order parameters in real space, e.g.
for the spin-$\frac{1}{2}$ case, it takes the following form:
\begin{eqnarray}
  \psi_{\bar{n},m}(\bm{\rho})&\propto&
  \int_0^{\infty}k'^{1/2}e^{-(k'-v')^2}H_{\bar{n}}(k'-v')\nonumber\\
  &&\times\left(
  \begin{array}{c}
    i^{m}J_m(k\rho)e^{im\varphi} \\
    -i^{m+1}J_{m+1}(k\rho)e^{i(m+1)\varphi}\\
  \end{array}
    \right) dk,
    \label{spinhalfv}
\end{eqnarray}
where $\varphi=\arg(x+iy)$ and $J_m$ is the Bessel function of the first kind.
All states given in Eq. (\ref{eigenstates}) possess the continous gauge, spin, and space rotation symmetries. 
Some of them can be stabilized with weak interactions.
For instance, two different vortices are found in the pseudo spin-$\frac{1}{2}$ case,
and their order parameters take the form given in Eq. (\ref{spinhalfv}) with $\bar{n}=0$ and $m=-2,-1,0,1$.
They are invariant under $e^{i\kappa\theta}\mathcal{C}_z(\theta)$ with $\kappa=\pm1/2$ (see Fig. 3(d) of Ref. \cite{hu2012}
and Fig. 1(a) of Ref. \cite{sinha2011}) and $\kappa=\pm3/2$ (see Fig. 3(b) of Ref. \cite{hu2012}
and Fig. 1(b) of Ref. \cite{sinha2011}). 
From the combined symmetry, we know that the vortex-core state at the rotating center should 
be invariant under the spin-gauge symmetry $e^{i\kappa\theta}e^{-iF_z\theta}$.
Therefore, we determine the vortex core as that summarzied in Table \ref{tabl}.

We would like to point out that such vortices with continous combined symmetry
also appear in a spinor dipolar BEC where the effective SO coupling arises from 
long-range magnetic dipole interaction. For instance, two different ground states
preserving continous combined symmetry are found in the ground state of the spin-1 $^{87}$Rb condensate \cite{yuki2006,yi2006}.
One called the chiral spin vortex phase is invarint under $e^{i\theta}\mathcal{C}_z(\theta)$, 
which results in a vortex with the vortex core filled by the ferromagnetic phase.
The other called the polar-core vortex phase preserves $\mathcal{C}_z(\theta)$,
which induces a polar-core vortex because $e^{-iF_z\theta}|M_F=0\rangle=|M_F=0\rangle$.

\subsection{Discrete combined symmetry}
According to the symmetry classification scheme \cite{xu2012},
there are states preserving discrete rotation symmetries. Some of them
spontaneously emerge as the ground states with individual spin-component density
distributions showing lattice structures \cite{xu2011,kawakami2011,hu2012,sinha2011,xu2012}.
In the strong SO-coupling limit, the order parameter can be well approximated
by a superposition of several plane waves given in Eq. (6) of Ref. \cite{xu2012} as
\begin{eqnarray}
  \psi= \sum\limits_{j=0}^{\mathbb{N}-1}e^{i\phi_j}e^{i\mathbf{b}_{j}\cdot\bm{\rho}}\zeta_{-F}(j2\pi/\mathbb{N}),
  \label{latticephases}
\end{eqnarray}
where $\mathbf{b}_j=k_g(\cos(j2\pi/\mathbb{N}),\sin(j2\pi/\mathbb{N}))$
and $k_g=mv/\hbar$. 

\subsubsection{spin-1/2}
Firstly, we focus on the pseudo spin-$\frac{1}{2}$ case,
where three different types of lattice phases are found in the ground state.
Figure \ref{fig2} illustrates their spin textures which are obtained by numerically solving
the coupled Gross-Pitaevskii equations
with $v'=15$ and (a) $\alpha=0.5\hbar\omega_{\perp}/N$ and $\beta=-0.1\alpha$, 
(e) $\alpha=0.3\hbar\omega_{\perp}/N$ and $\beta=0.1\alpha$, and (g)
$\alpha=0.4\hbar\omega_{\perp}/N$ and $\beta=0.1\alpha$, 
and compares them with 
those of the variational order parameters described in Eq.~(\ref{latticephases})
with parameters
(b) $\mathbb{N}=3$ and $(\phi_0,\phi_1,\phi_2)=(\pi/3,\pi,5\pi/3)$,
(f) $\mathbb{N}=6$ and $(\phi_0,\phi_1,\phi_2,\phi_3,\phi_4,\phi_5)=(\pi/6, \pi, 5\pi/6, -\pi/3, -\pi/2, \pi/3)$, and
(h) $\mathbb{N}=6$ and $\phi_j=(j+1)\pi/3$ with $j=0, \dots, 5$.
The validity of the ansatz in Eq.~(\ref{latticephases}) is numerically confirmed.
In the following, we discuss the symmetry of the vortices based
on the ansatz. Strictly, the symmetry we discuss below can only 
be approximately satisfied. If we increase the strength of SO coupling,
the agreement between the ansatz and numerically exact result becomes better.

The triangular-lattice phase shown in Figs.~\ref{fig2}(a-d)
is found to be composed of half and anti-half skyrmions (enclosed in the triangle $A_0A_1A_2$
and the one nearby), which appear alternatively
and are connected by singular vortices (with one locating at $A_0$).
Based on the symmetry analysis in Sec.~\ref{symmetry}, we find that for the case of $n=3$ there are three different types of symmetries
generated by $e^{i\kappa 2\pi/3}\mathcal{C}_z(2\pi/3)$ with $\kappa=1/2, 3/2, 5/2$, respectively.
As presented in Ref. \cite{xu2012}, they are connected by shifting the lattice
as shown in Figs.~\ref{fig2}(a-d), where three different vortices preserving
different combined discrete symmetries are denoted by $A_0$, $A_3$ and $A_4$,
filled by the vacuum, spin-up, and spin-down states, respectively.
To confirm this statement, we focus on the order parameter 
\begin{eqnarray}
  \psi&=&e^{i\pi/3}e^{ik_gx}\zeta_{-\frac{1}{2}}(0)
  +e^{i\pi}e^{ik_g(-\frac{x}{2}+\frac{\sqrt{3}y}{2})}\zeta_{-\frac{1}{2}}(2\pi/3)
  \nonumber\\
  &&+e^{i5\pi/3}e^{ik_g(-\frac{x}{2}-\frac{\sqrt{3}y}{2})}\zeta_{-\frac{1}{2}}(4\pi/3),
  \label{triangular}
\end{eqnarray}
which is shown in Fig.~\ref{fig2}(b).
It is invaraint under $e^{i\pi}\mathcal{C}_z(2\pi/3)$.
Therefore, vortices at the rotating center denoted by $A_0$ is singular.
Shifting the lattice center to the point $A_3$ by coordinate 
transformation $\{x+2\pi/3k_g\rightarrow x,~y-2\pi/3\sqrt{3}k_g\rightarrow y\}$,
the order parameter is transformed into 
\begin{eqnarray}
  \psi&=&e^{i5\pi/3}e^{ik_gx}\zeta_{-\frac{1}{2}}(0)
  +e^{i5\pi/3}e^{ik_g(-\frac{x}{2}+\frac{\sqrt{3}y}{2})}\zeta_{-\frac{1}{2}}(2\pi/3)
  \nonumber\\
  &&+e^{i5\pi/3}e^{ik_g(-\frac{x}{2}-\frac{\sqrt{3}y}{2})}\zeta_{-\frac{1}{2}}(4\pi/3),
  \label{triangularf}
\end{eqnarray}
which is invariant under $e^{i\pi/3}\mathcal{C}_z(2\pi/3)$. We thus find that the vortex
core is filled by the spin-up state. Similar arguments can be applied to vortices located
at $A_1$, $A_2$ and $A_4$.

In constrast to the triangular-lattice phase, two kagome-lattice phases shown
in Figs.~\ref{fig2}(e,f) and (g,h) cannot be recast into half- and anti-half-skyrmion lattices.
They possess combined discrete symmetries with symmetry groups generated
by $\{e^{i(2\pi/3+\pi/3)}\mathcal{C}_z(2\pi/3), \mathcal{T}\mathcal{C}_z(\pi)\}$ and
$\{e^{i(\pi/3+\pi/6)}\mathcal{C}_z(\pi/3)\}$ \cite{note}, respectively.
Therefore, we expect that the vortex in the rotating center should be filled by 
the vacuum state, as denoted by $B_0$ and $C_0$, respectively.
Besides, there are two more different singular vortices denoted by $B_1 (C_1)$ and $B_2 (C_2)$,
originating from different symmetries, which can be determined by shifting the 
lattice center as that applied in discussing vortices in the triangular-lattice phase.

Refocusing on spin textures shown in Fig.~\ref{fig2},
we do see that for vortices at $A_0$, $B_0$, $B_1$, $B_2$, $C_1$ and $C_2$, the total spin rotation around the vortex core 
is $-4\pi$, different from $2\pi$. The corresponding quantized number $\mathbb{N}_s$ is $-1$. Such vortices are unique
to the discrete combined rotation symmetry.
By examining the phase winding number for each spin component, we can determine another quantum number
$\mathbb{N}_g$ by solving Eq.~(\ref{windingnumber2}).
To show this explicitly, we take the vortex at $A_0$ preserving $e^{i\pi}\mathcal{C}_z(2\pi/3)$ as an example.
From Figs.~\ref{fig2}(c-d), we infer that the corresponding phase winding numbers for two spin components are
$w_{\frac{1}{2}}=1$ and $w_{-\frac{1}{2}}=-1$. Using Eq.~(\ref{windingnumber2}) with $M_F=\frac{1}{2}$, we get
\begin{eqnarray}
  \frac{3}{2}+3\mathbb{N}_g-(1-3)\left(\frac{1}{2}\right)=1.
\end{eqnarray}
Therefore, we find $\mathbb{N}_g=-\frac{1}{2}$.

\begin{figure*}[t]
\centering
\includegraphics[width=5.7in]{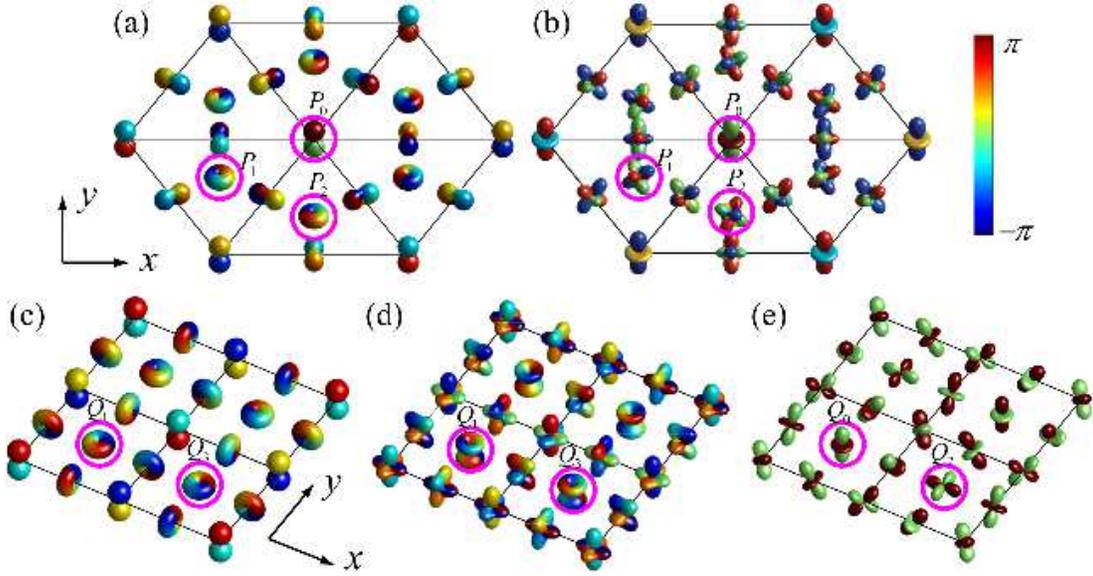}
\caption{(color online). Spin textures of the triangular-lattice phase (a, b) and the square-lattice phase (c, d, e)
of spin-orbit-coupled spin-1 (a, c) and spin-2 (b, d, e) BECs. Each subfigure shows spatial variation of the corresponding order parameter
in a unit cell visualized by plotting $\sum_{m}\xi_{m}Y_F^{m}(\theta,\varphi)$ with its phase displayed according to the right color gauge,
where $Y_F^{M_F}$ is the spherical harmonics of rank $F$ and $\xi_m$ is the $m$-th component
of the spin wave function. For the triangular-lattice phase, there are vortices with cores at the vertices and the center of the triangles.
For the square-lattice phase, vortex cores are located at the center of the squares. 
One representative vortex for each type of symmetry described by $e^{i\kappa 2\pi/3}\mathcal{C}_z(2\pi/3)$ [$e^{i\kappa\pi/2}\mathcal{C}_z(\pi/2)$] is enclosed
in the circle `$\circ$' and denoted by $P_{\kappa}$ [$Q_{\kappa}$].  }
\label{fig3}
\end{figure*}

\subsubsection{spin-1 $\&$ spin-2}
Next, we focus on the case of integer spin, where we find triangular and square-lattice phases.
They can all be well described by superposition
of three or four plane waves as described in Ref.~\cite{xu2012} and in Eq.~(\ref{latticephases}) with $\mathbb{N}=3$ and $4$.
Similar to the spin-$\frac{1}{2}$ case,
the combined symmetry is approximately satisfied for numerically derived ground states.
In the following, we only use their corresponding variational order parameters in Eq.~(\ref{latticephases}) to discuss the symmetry.
Figure 2 shows the corresponding spin textures in a unit cell of five lattice phases found in 
SO-coupled spin-1 and spin-2 BECs. For the triangular-lattice phase, we can choose arbitrary values
of $\phi_j$ in Eq.~(\ref{latticephases}), as they are connected by shifting the lattice center.
For the square-lattice phase in Figs.~\ref{fig3}(c-e), one representative state in Eq.~(\ref{latticephases})
can be chosen as (c, d) $\phi_j=j\pi/2$  and (e) $\phi_j=0$ with $j=0,\dots,3$.

Similar to the pseudo spin-$\frac{1}{2}$ case, three different vortices appear simultaneously
in a triangular-lattice phase due to three different combined discrete symmetries
generated by $e^{i\kappa 2\pi/3}\mathcal{C}_z(2\pi/3)$ with $\kappa=0, 1, 2$, respectively.
They are connected by shifting the lattice center \cite{xu2012}.
By directly solving Eq. (\ref{spingauge}), we uniquely determine the vortex-core spin state 
for the spin-1 case. This breaks down for the spin-2 case on two specific symmetries of
$e^{i2\pi/3}\mathcal{C}_z(2\pi/3)$ and $e^{i4\pi/3}\mathcal{C}_z(2\pi/3)$, as there are infinite vortex-core
spin states satisfying the same combined symmetry. For instance, the spin state $c_1|M_F=1\rangle+c_2|M_F=-2\rangle$
with arbitrary coefficients $c_1$ and $c_2$ is invariant under $e^{i2\pi/3}e^{-iF_z2\pi/3}$, 
and the one which is actually chosen is determined by the Hamiltonian. 
In Figs.~\ref{fig3}(a) and (b), we indicate by circles three different types of vortices which preserve
different symmetries.

For the case of $n=4$, as discussed in Sec.~\ref{ourmethod}, there are four different vortices preserving
the combined discrete symmetries generated by $e^{i\kappa\pi/2}\mathcal{C}_z(\pi/2)$ with $\kappa=0,1,2,3$.
They are found in two different square-lattice phases
with one breaking time-reversal symmetry as illustrated in Figs.~\ref{fig3}(c) and (d)
and the other one preserving time-reversal symmetry as illustarated in Fig.~\ref{fig3}(e).
As shown in Ref. \cite{xu2012}, in a time-reversal symmetry breaking square-lattice phase,
the state can preserve two different combined symmetries of $e^{i\pi/2}\mathcal{C}_z(\pi/2)$ and 
$e^{i3\pi/2}\mathcal{C}_z(\pi/2)$ by choosing the proper center of the lattice.
This results in two different vortices with the corresponding vortex-core spin state
preserving the spin-gauge symmetry $e^{i\pi/2}e^{-iF_z\pi/2}$ and $e^{i3\pi/2}e^{-iF_z\pi/2}$.
Solving Eq. (\ref{spingauge}), we can uniquely determine the vortex-core
spin state, as demonstrated in Figs.~\ref{fig3}(c) and (d).
The other type of square-lattice phase, which preserves time-reversal symmetry
and $\mathcal{C}_z(\pi/2)$ or $e^{i\pi}\mathcal{C}_z(\pi/2)$ relying on the choice of the lattice center,
has only been found in the spin-2 case. These symmetries induce two different
nonsingular vortices with vortex cores determined from Eq. (\ref{spingauge})
and the time-reversal symmetry requirement. These two vortices are shown in Fig. \ref{fig3}(e).

\begin{figure}[tpb]
\centering
\includegraphics[width=\columnwidth]{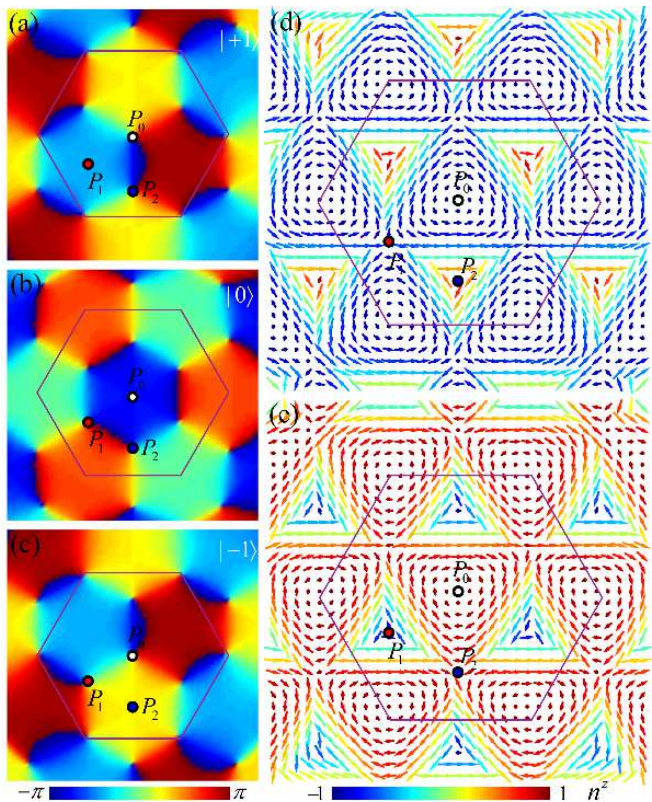}
\caption{(color online). Phase distributions for individual spin components ($M_F=1,0,-1$) (a-c) and 
  the corresponding spin textures for two unit vectors using the Majorana representation (d-e) of the ground state 
  showing triangular structure of a spin-orbit-coupled pseudo-spin-1 BEC with $v'=15$, $\alpha=0.5\hbar\omega_{\perp}/N$ and $\beta=0.2\alpha$, 
A unit cell of the order parameter is indicated by a solid hexagon. Vortices denoted by $P_0$, $P_1$ and $P_2$
are the same as those in Fig.~\ref{fig3}(a).
}
\label{fig4}
\end{figure}

Reconsidering vortices shown in Fig.~\ref{fig3}, we find that there are new vortices which are
unique to the discrete combined symmetry
with at least one nonzero quantum number $\mathbb{N}_s^{(M_F)}$ related to the total spin rotation.
To demonstrate the searching procedure explicitly, we take the triangular-lattice phase of spin-1 BECs
as an example. The ground-state phase distributions for three spin components 
are shown in Figs.~\ref{fig4}(a-c). To examine the vortex at $P_1$ preserving $e^{i2\pi/3}\mathcal{C}_z(2\pi/3)$,
we use Eq.~(\ref{windingnumberF}) to derive the following three equations:
\begin{eqnarray}
  &1+3\mathbb{N}_g-\left(1+3\mathbb{N}_s^{(1)}\right)=0,&\nonumber\\
  &1+3\mathbb{N}_g-\left(1+3\mathbb{N}_s^{(-1)}\right)(-1)=-1,&\nonumber\\
  &1+3\mathbb{N}_g=1.&
\end{eqnarray}
Solving these equations, we find $(\mathbb{N}_g, \mathbb{N}_s^{(1)}, \mathbb{N}_s^{(-1)})=(0,0,-1)$.
Therefore, this vortex is topologically different from vortices preserving the continuous combined symmetry.
Applying the same method to vortices at $P_0$ and $P_2$ preserving $\mathcal{C}_z(2\pi/3)$ and $e^{i4\pi/3}\mathcal{C}_z(2\pi/3)$, 
respectively, we find their three quantum numbers as $(\mathbb{N}_g, \mathbb{N}_s^{(1)}, \mathbb{N}_s^{(-1)})=(0,0,0)$
and $(-1,-1,0)$.
 
As discussed in Sec.~\ref{ourmethod}, the reason why we need $2F$ quantized numbers to 
describe total spin rotation is that to denote each spin-$F$ state,
we need $2F$ unit vectors, which can be obtained by using 
the Majorana representation \cite{barnett2006,yuki2011},
where $2F$ unit vectors
$(n^x,n^y,n^z)=(\sin\theta\cos\varphi,\sin\theta\sin\varphi,\cos\theta)$
correspond to the $2F$ solutions of the equation
\begin{eqnarray}
  b_0+b_1\xi+\dots+b_{2F}\xi^{2F}=0,
  \label{majorana}
\end{eqnarray}
where $\xi=e^{i\varphi}\tan(\theta/2)$
and $b_i=\psi^*_{F-i}/\sqrt{(2F-i)!i!}$.
Figures \ref{fig4}(d-e) show the corresponding spin textures of two unit vectors for the ground state in Figs.~\ref{fig4}(a-c),
where we classify two unit vectors numerically by comparing values of their $z$-component $n^z$.
We can infer that both two vectors rotate around the rotating axis at $P_0$ for $2\pi$,
whereas for vortices at $P_1$ and $P_2$, the total rotation for one vector is $-4\pi$.
This is consistent with the above arguments that the vortex at $P_0$ is topologically equivalent to 
vortices which preserve the continuous combined symmetry, whereas vortices at $P_1$ and $P_2$ are 
topologically different.

\subsection{Lattice translation symmetry}
\begin{figure*}[t]
\centering
\includegraphics[width=5.0in]{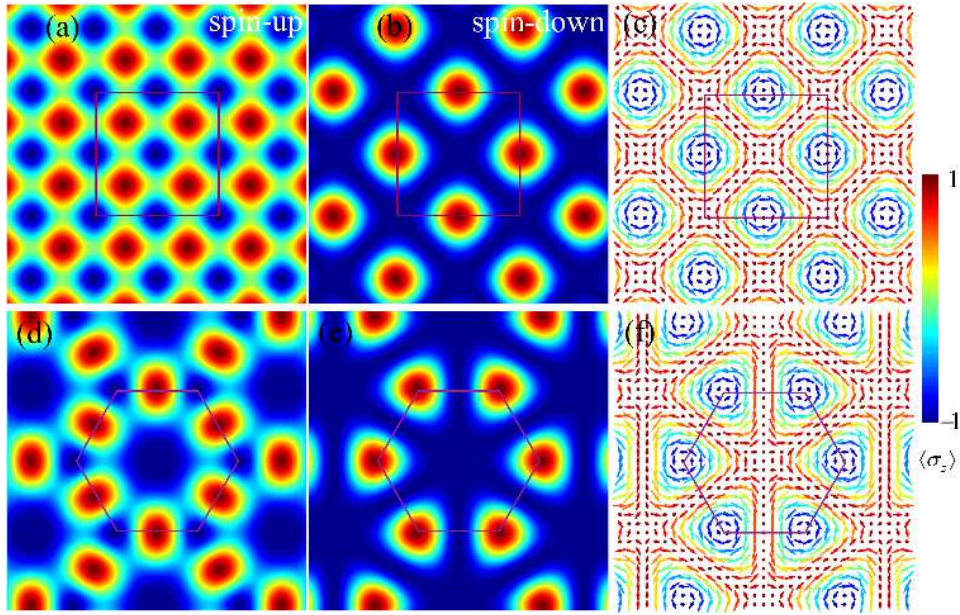}
\caption{(color online). Two different skyrmion crystals predicted from the symmetry classfication for the spin-$\frac{1}{2}$
case \cite{xu2012}, where their spin-up, spin-down component density distributions, and spin textures 
are shown from left to right.
The corresponding order parameters are given by Eq.~(\ref{latticephases})
with (a-c) $\mathbb{N}=4$ and $\phi_j=j\pi/2$ ($j=0,\dots,3$) and (d-f) $\mathbb{N}=6$
and $\phi_j=j2\pi/3$ ($j=0,\dots,5$). A unit cell is indicated by (c) a solid square and (f) a solid hexagon,
both with $Q=-2$.  }
\label{fig5}
\end{figure*}

For lattice phases, vortices form lattice structures. Besides preserving the discrete combined rotation symmetry, 
their order parameters are approximately invariant under lattice translation.
For such lattice structures we can define a topological charge
for each unit cell which is analogous to the Chern number:
\begin{eqnarray}
  \mathcal{C}_h=\iint_{\rm uc}(\nabla\times\mathbf{v}^{\rm mass})_z dxdy,
  \label{chernnumber}
\end{eqnarray}
where the superfluid velocity $\mathbf{v}^{\rm mass}$ is given by
\begin{eqnarray}
  \mathbf{v}^{\rm mass}=\frac{\hbar}{2M|\psi|^2i}[\psi^*\nabla\psi-(\nabla\psi)^*\psi].
  \label{velocity}
\end{eqnarray}
For the integer spin cases, the topological charge for different lattice phases
in the ground state vanishes, as there are only nonsingular vortices and the order parameter changes smoothly over the entire manifold.
For the pseudo spin-$\frac{1}{2}$ case, we have confirmed that all three different lattice phases have zero topological charges,
although there are singular vortices. Based on the Mermin-Ho relation \cite{mermin1976}
\begin{eqnarray}
  \left(\nabla\times\mathbf{v}^{\rm mass}\right)_{\mu}
  =\frac{\hbar}{4\pi M}\epsilon_{\mu\nu\tau}\mathbf{S}\cdot(\partial_{\nu}\mathbf{S}
  \times\partial_{\tau}\mathbf{S}),
  \label{merminho}
\end{eqnarray}
where $S_{\mu}=\langle\sigma_{\mu}\rangle$ and $(\mu, \nu, \tau)=x,y,z$,
we find that the total skyrmion number $Q_{\rm SkX}$ for each unit cell is zero, where 
\begin{eqnarray}
  Q_{\rm SkX}=\frac{1}{4\pi}\iint_{\rm uc} \mathbf{S}\cdot(\partial_x\mathbf{S}\times\partial_y\mathbf{S}) dx dy.
  \label{skyrmion}
\end{eqnarray}
Therefore, the lattice phases found in the ground state of a SO-coupled BEC
are quite different from skyrmion crystals (SkXs) with a nonzero topological charge for each unit cell found in chiral magnets \cite{muhlbauer2009},
where the Dzyaloshinskii-Moriya (DM) interaction \cite{han2012} plays a crucial role.
However, we find that for the spin-$\frac{1}{2}$ case, several lattice phases, which we predict from the symmetry classification \cite{xu2012} 
but have not been found in the ground states, show similar structures as SkXs.
Figure~\ref{fig5} shows two different SkXs, where
the corresponding order parameters are given by Eq.~(\ref{latticephases})
with (a-c) $\mathbb{N}=4$ and $\phi_j=j\pi/2$ ($j=0,\dots,3$) and (d-f) $\mathbb{N}=6$
and $\phi_j=j2\pi/3$ ($j=0,\dots,5$).
From the spin textures in Fig.~\ref{fig5}(c) and (f), we can infer 
that both SkXs have the same topological charge $Q_{\rm SkX}=-2$.
Here, the unit cell is determined from the periodicity of the order parameter.
By performing time reversal, we obtain two other SkXs with $Q_{\rm SkX}=2$.
The square shape of the SkX in Fig.~\ref{fig5}(a-c) has also been found as a stationary state in a system without SO coupling \cite{zhang2013}.

\section{Vortices in spinor BECs}
In this section, we generalize the symmetry classification of vortices to a system in which the spin and orbital
degrees of freedom are decoupled. Examples include spinor BECs and rotating spinor BECs, where the corresponding Hamiltonian is invariant
under the U(1) gauge transformation, SO(3) spin rotation, and SO(2) space rotation about the $z$-axis.
For spinor BECs, we can choose their Hamiltonian as a special case in Sec.~\ref{hamiltonian} with $v=0$.
Furthermore, by adding a rotating term $\mathcal{H}_{\rm rot}=-\Omega \int d\bm{\rho}\ \hat{\psi}^{\dag}L_z\hat{\psi}$, where $\Omega$ and $L_z$ refer
to the rotating frequency and the projected orbital angular momentum, we obtain the Hamitlonian to describe a rotating spinor BEC.

As long as the spin and orbital degrees of freedom are decoupled,
when spontaneous symmetry breaking occurs, there are other types 
of vortices which are invariant under the combined symmetry described by $e^{i\phi}e^{-i\theta F_z}\mathcal{R}_z(\vartheta)$,
where $\phi$, $\theta$, and $\vartheta$ refer to the gauge, spin, and space rotation angles, respectively,
and $\mathcal{R}_z(\vartheta)$ describes the $\theta$ space rotation about the $z$-axis,
because $\theta$ and $\vartheta$ can take different values.
As there are numerous vortices from the viewpoint of the spontaneous symmetry breaking,
in the following, we focus on a special vortex with fixed boundary conditions and predict their possible
vortex cores based on the combined symmetry in a manner similar to the classification scheme of 
vortices in superfluid $^{3}$He-B \cite{salomaa1983,salomaa1986}.

For instance, we take a fractional $1/2$-$1/4$ vortex in the biaxial nematic phase of spin-2 BECs discussed in Ref. \cite{kobayashi2009},
where $1/2$ and $1/4$ refer to the $\pi$ gauge transformation and the $\pi/2$ spin rotation around the vortex, respectively.
Therefore, away from the vortex core, the axisymmetric order parameter can be chosen as 
$\psi(\rho\rightarrow\infty)=e^{i\varphi/2}e^{-i\varphi F_z/4}\psi_B=\sqrt{\rho_s}(1,0,0,0,e^{i\varphi})^T/\sqrt{2}$,
where $\psi_B=\sqrt{\rho_s}(1,0,0,0,1)^T/\sqrt{2}$ is the order parameter for the biaxial nematic phase, and $\varphi$ is the azimuthal angle.

Under this boundary condition, the continuous combined symmetry which can be satisfied by the vortex
is described by operators $e^{i2\theta}e^{-i\theta F_z}\mathcal{R}_z(4\theta)$ with $\theta\in[0,\pi/2]$,
which implies that the $\pi$ gauge transformation and the $\pi/2$ spin rotation are uniformly distributed along 
the azimuthal direction. When $\theta=\pi/2$, we get $e^{i\pi}e^{-i F_z\pi/2}\mathcal{R}_z(2\pi)|\psi\rangle=e^{i\pi}e^{-i F_z\pi/2}|\psi\rangle=|\psi\rangle$,
which results in a strong constraint
for the order parameter of the fractional $1/2$-$1/4$ vortex satisfying 
the continuous combined rotation symmetry, because $\psi$ should be a superposition of $|M_F=2\rangle$ and $|M_F=-2\rangle$ spin states 
everywhere. Meanwhile, we can determine the nonsingular vortex-core spin state at the rotating center
from the spin-gauge symmetry, and find that it is a ferromagnetic phase
with order parameter $\psi(\rho=0)=\sqrt{\rho_s}(1,0,0,0,0)^T$.
This type of fractional vortices filled by the ferromagnetic core has already been numerically found
in Ref. \cite{kobayashi2009}.

Besides the continuous combined symmetry, we can also search for a fractional $1/2$-$1/4$ vortex with a discrete combined symmetry,
which is described by the operator $e^{i\phi}e^{-i\theta F_z}\mathcal{R}_z(2\pi/n)$.
Here, we focus on the discrete symmetries with a constraint (A) $e^{in\phi}e^{-in\theta F_z}=\mathbb{1}_{2F+1}$.
Therefore, the order parameter is no longer fixed into a superposition of two spin states.
To be consistent with the boundary condition, there is another constraint (B)
$e^{i\phi}e^{-i\theta F_z}(1,0,0,0,0)^T=e^{i\pi/n}e^{-iF_z\pi/2n}(1,0,0,0,0)^T$.
By solving these two equations (A) and (B), we can predict possible types of fractional $1/2$-$1/4$
vortices distinguished by the discrete combined gauge, spin, and space rotation symmetry.
One solution is that $\phi=4\pi/3$, $\theta=2\pi/3$ and $n=3$.
Furthermore, we determine the nonsingular vortex-core spin state at the rotating center to be an arbitrary superposition of 
$|M_F=2\rangle$ and $|M_F=-2\rangle$ spin states. The coefficients of the superposition
depend on the Hamiltoinan. In Ref. \cite{kobayashi2009}, a specific vortex with a cyclic core is discussed.

We have predicted possible vortices and the corresponding vortex cores 
under the axisymmetric boundary condition for a fractional $1/2$-$1/4$ vortex.
Such symmetry classification is also applicable to understand other vortices
with axisymmetric boundary conditions.

\section{Conclusions}
We have systematically studied the structure of ground states of SO-coupled spinor BECs, including pseudo spin-$\frac{1}{2}$,
spin-1, and spin-2 systems. 
In constrast to quantized vortices in spinor BECs which are distinguished by toplogically
invariant quantized numbers, vortices in SO-coupled BECs spontaneously emerge in the groud states,
and are classified by the combined gauge, spin, and space rotation symmetry and even
time reversal. In addition, the symmetry classification not only classifies vortices 
but also determines the vortex-core spin state. We have studied vortices with both continuous and discrete symmetries.
For vortices preserving the continuous combined symmetry, the total spin rotation around the vortex
core is fixed to be $2\pi$, whereas for vortices invariant under the discrete combined symmetry,
we need a further quantum number $\mathbb{N}_s$ for the spin-$\frac{1}{2}$ case and $2F$ quantum numbers $\mathbb{N}_s^{(M_F)}$ 
with $M_F=\pm1,\dots,\pm F$ for the spin-$F$ case to specify the total spin rotations.
From numerical results, we do find new vortices with nontrivial quantized numbers ($\mathbb{N}_s\ne0$,
$\mathbb{N}_s^{(M_F)}\ne0$).
For lattice phases, where the order parameter preserves lattice translation symmetry,
we define a Chern-number-like topological charge for each unit cell.
Unfortunately, all lattice phases found in the ground state have zero topological charge.
However, we find two types of skyrmion crystals in the BEC phase of the spin-$\frac{1}{2}$ system.
They are highly symmetric states predicted from the symmetry classification, and may be experimentally realized as stationary or excitated states.
Finally, we also show that the symmetry classification of vortices is also applicable to other systems
such as spinor dipolar BECs, spinor BECs and rotating spinor BECs.

\section*{ACKNOWLEDGEMENTS}
ZFX is grateful to Yuki Kawaguchi and Jia-Wei Mei for valuable discussions. 
This work is supported by Grants-in-Aid for Scientific Research (KAKENHI 22340114 and 22103005),
a Global COE Program ``the Physical Sciences Frontier'',
the Photon Frontier Network Program, from MEXT of Japan, and NSFC (No.~11004116). 
ZFX acknowledges support from JSPS (Grant No. 2301327).
SK acknowledges support from JSPS (Grant No. 228338).

\end{document}